\documentclass[12pt,A4paper]{article}
\usepackage{graphicx}
\usepackage{setspace}
\usepackage{color}
\usepackage{amsmath}
\usepackage{amssymb}
\usepackage{cancel}
\usepackage{url}

\date{}
\begin{document} 

\title{Attosecond metrology of 2D charge distribution in molecules} 
\maketitle

\author{V. Loriot$^{1,*}$, A. Boyer$^{1}$, S. Nandi$^{1}$, C. M. Gonz\'alez-Collado$^{2}$, E. Pl\'esiat$^{3}$, A. Marciniak$^{1}$, C. L. Garcia$^{1}$, Y. Hu$^{1}$, M. Lara-Astiaso$^{2}$, A. Palacios$^{2,4,6}$, P.~Decleva$^{5}$, F. Mart\'in$^{2,3,6}$ and F. L\'epine$^{1,*}$}\\
\newline
\normalsize{$^{1}$Univ Lyon, Univ Claude Bernard Lyon 1, CNRS, Institut Lumi\`ere Mati\`ere, F-69622, VILLEURBANNE, France}\\
\normalsize{$^{2}$Departamento de Qu\'imica, M\'odulo 13, Universidad Aut\'onoma de Madrid, 28049 Madrid, Spain}\\
\normalsize{$^{3}$Instituto Madrile\~{n}o de Estudios Avanzados en Nanociencia (IMDEA-Nanociencia), Cantoblanco, 28049 Madrid, Spain}\\
\normalsize{$^{4}$Institute for Advanced Research in Chemical Sciences (IAdChem), Universidad Aut\'onoma de Madrid, 28049 Madrid, Spain} \\
\normalsize{$^{5}$Dipartimento di Scienze Chimiche e Farmaceutiche, Universit\`a degli Studi di Trieste and CNR-IOM, 34127 Trieste, Italy}\\
\normalsize{$^{6}$Condensed Matter Physics Center (IFIMAC), Universidad Aut\'onoma de Madrid, 28049 Madrid, Spain}\\
\\
\normalsize{$^\ast$ Corresponding authors :} \\
\normalsize{E-mail: vincent.loriot@univ-lyon1.fr and franck.lepine@univ-lyon1.fr.}

\newpage
\abstract{
\textbf{
Photoionization as a half-scattering process is not instantaneous. Usually, time delays in photoionization are of the order of few tens of attoseconds (1 as = 10$^{-18}$~s). While going from a single atom to a nano-object, one can expect the delay to increase since the photoelectron scatters over a larger distance. Here, we show that this is no longer valid when comparing three dimensional and planar systems. Using attosecond interferometry, we find that the time delays in 2D carbon-based molecules can significantly be smaller than those of the corresponding 3D counterparts, for example, naphthalene compared to adamantane. The measured time delay carries the signature of the spatial distribution of the 2D hole created in the residual molecular cation, allowing us to obtain its dimensions with angstrom accuracy. Our findings offer novel opportunities for tracking and manipulating ultrafast charge transport in molecular materials.
} 
}

\vspace{1cm}

Electrons are very accurate probes of matter at the fundamental level. In electron microscopy, their quantum nature and short De Broglie wavelength permits to reach sub-nanometer resolution by measuring the properties of the transmitted and scattered electrons from a sample \cite{Hawkes2007}. In photoionization spectroscopy, measuring the electron momentum allows us to determine the electronic structure of bulk matter, molecules and atoms \cite{becker}. In this context, attosecond technology \cite{RMPAttoworld} has opened up a wide variety of novel opportunities to study dynamical properties of matter on the angstrom length scale \cite{lepineNatPhoton}. In attosecond science, a coherent ultrashort extreme ultraviolet (XUV) pulse is often used as a precise stopwatch for the photoionization step \cite{RMP2015}. During the photoionization process, the ejected photoelectron is dispersed by the short range potential close to the ionic core of an atom or a molecule. In terms of quantum scattering, the local interaction with the atomic or molecular environment is imprinted on the phase of the photoelectron wavepacket. The derivative of this phase with respect to kinetic energy corresponds to the Wigner delay: the time it takes for the particle to scatter from the potential \cite{WignerIonizTime}. Nowadays, such delays in photoionization have become accessible by using interferometric approaches in which XUV attosecond pulses and a near infrared (NIR) pulse are recombined with attosecond precision, producing photoelectrons that carry information on the scattering process. Attosecond photoionization delays have been measured in atoms \cite{Schultze1658,klunder2011,ScienceLHuillier2017}, and small molecules \cite{PRLWornerH2ON2O,VosScience2017,biswas2020,Loriot2020,Nandi2020}, and recently in water clusters \cite{WaterClusterWorner2022} unraveling the influence of angular momentum, electron correlation, chemical environment and resonances. The delays induced by the inelastic scattering in dielectric nano-particles were measured by Seiffert \textit{et al.} \cite{seiffert2017}, while in solids, Ossiander \textit{et al.} \cite{Ossiander2018} measured the absolute exit delay associated with an electron escaping the core level in tungsten crystals. Recently, photoionization delays have been used to study the non-local character of ionization in liquid phase \cite{WornerScienceH2O_2020} but so far, the effect of the spatial extension encountered in large molecules remains unexplored.

Here, we have used the RABBIT (Reconstruction of Attosecond Beating By Interference of Two-photon transitions) technique \cite{Paul1689} to measure the photoionization delays in  two-dimensional (2D) hydrocarbons, namely, naphthalene (C$_{10}$H$_{8}$), pyrene (C$_{16}$H$_{10}$) and fluorene (C$_{13}$H$_{10}$) and in the three-dimensional (3D) diamond-like molecule adamantane (C$_{10}$H$_{16}$). These molecules have been chosen as model, well-documented, systems to test our method. We have measured time-delay differences of several tens of attoseconds between 2D and 3D molecules, with the photoelectrons originating from 2D molecules being less delayed compared to those from adamantane. This significant difference in time delays can be attributed to the 2D spatial distribution of the hole in the residual molecular cation at the moment of ionization. These findings are supported by first-principles calculations based on static-exchange density functional theory (SE-DFT) that explicitly describes the electronic continuum of thoses molecules with proper scattering boundary conditions, as well as by an analytical model based on the first-order Born approximation.

\begin{figure*}[!htb]
\includegraphics[width=\textwidth]{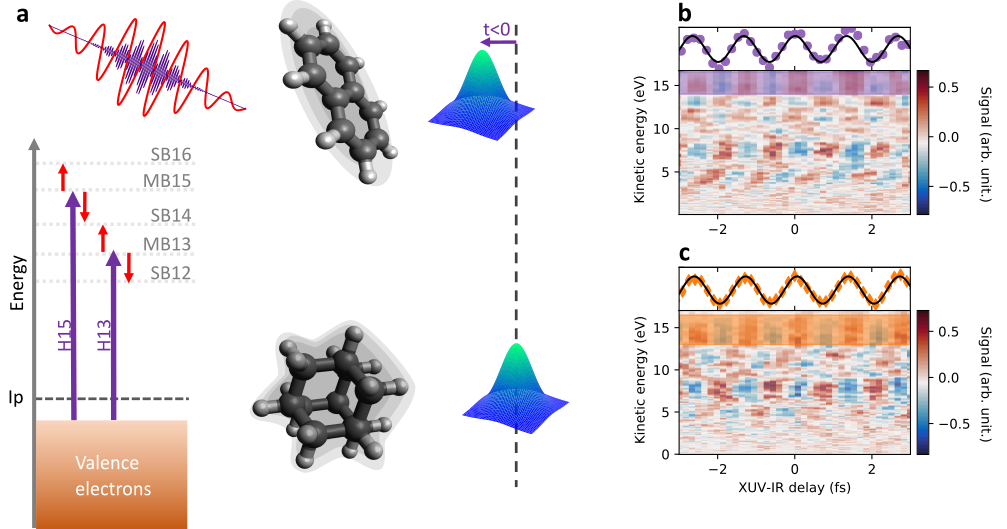}
\caption{{\bf Attosecond interferometry for delay measurements in complex molecules.} {\bf (a)} An APT ionizes the naphthalene (top) and the adamantane (bottom) molecules. In the spectral domain, the NIR dressing field couples the contributions from the APT harmonic comb allowing for interferences of the photoelectron wavepackets. 
{\bf (b)} and {\bf (c)} The measured RABBIT spectrograms for naphthalene and adamantane, respectively, from which the mean value of the signal has been subtracted. 
Sideband oscillations are only shown for the highest photoelectron kinetic energies, corresponding to the regions highlighted in the RABBIT maps. Compared to adamantane, photoelectrons from naphthalene are delayed by several tens of attoseconds. 
\label{Fig1}}
\end{figure*}

As shown in Fig.~\ref{Fig1}(a), the RABBIT protocol involves an attosecond pulse train (APT) produced via high-order harmonic generation (HHG), which is then spatio-temporally overlapped with the fundamental near-infrared (NIR) pulse (see Methods for details). In the spectral domain, the APT represents a comb of odd-order harmonics. Ionization of an electronic state by a photon from these harmonics leads to photoelectrons in the main-band. A NIR dressing field couples the ionization channels with each other leading to peaks (called sideband (SB)) between the main-bands (MB) (see, Fig.~\ref{Fig1}(a)). The sideband intensity, $I_{SB}(t)$, oscillates at twice the NIR optical frequency ($\pi/\omega_0=1.33$~fs) when varying the delay $t$ between the two pulses. The phases of the oscillations ($\phi$) depend on the phases due to the two photon (XUV-IR) transition ($\phi_{2h\nu}$) and the phases of the harmonics ($\phi_{HH}$), known as the `attochirp' \cite{mairesseScience2003}. Combining both, we can write,
\begin{equation}
    I_{SB}(t)\propto \mathcal{A} + \mathcal{B}\cos\left(2\omega_0t- \phi\right) = \mathcal{A} + \mathcal{B}\cos\left[2\omega_0t- \left(\Delta \phi_{2h\nu} + \Delta \phi_{HH}\right)\right].
    \label{RabbitOscillation}
\end{equation}
The corresponding RABBIT spectrograms for naphthalene and adamantane are shown in Fig.~\ref{Fig1}(b) and (c). We have used the oscillations related to the highest electron kinetic energy regions (see the highlighted parts of the RABBIT maps), which are associated with photoelectron emission mainly from the highest occupied molecular orbital (HOMO). In the present case, we calibrated the experimental $\phi$ values using an atomic target as a reference \cite{Ossiander2018}. For a given harmonic comb, the attochirp is identical for photoelectrons originating from two different systems. Any difference in $\phi$ between the two molecules can therefore be considered as a pure molecular effect arising from different two-photon phases. In the absence of narrow resonances in the ionization continuum, the two-photon phase $\Delta \phi_{2h\nu}$ can be split into two terms: the phase coming from the XUV molecular photoionization ($\Delta \phi_{mol}$) and the continuum-continuum phase ($\Delta \phi_{cc}$) due to the NIR-assisted transition following the photoionization:  $\Delta \phi_{2h\nu}\approx \Delta \phi_{mol} + \Delta \phi_{cc}$. 
To a good approximation, $\Delta \phi_{cc}$ depends only on the kinetic energy of the electrons and the NIR frequency \cite{Dahlstrom2012}. As the ionization potentials of the two molecules are rather similar \cite{NIST2001}, the contributions from this particular term are also similar (see Methods and Fig.~\ref{phicc} of the Supplementary Information). Thus, the difference in experimental $\Delta \phi_{2h\nu}$ values can effectively be considered to emerge from the difference in the $\Delta \phi_{mol}$ values between the two molecules.

\begin{figure*}[!htb]
\includegraphics[width=\textwidth]{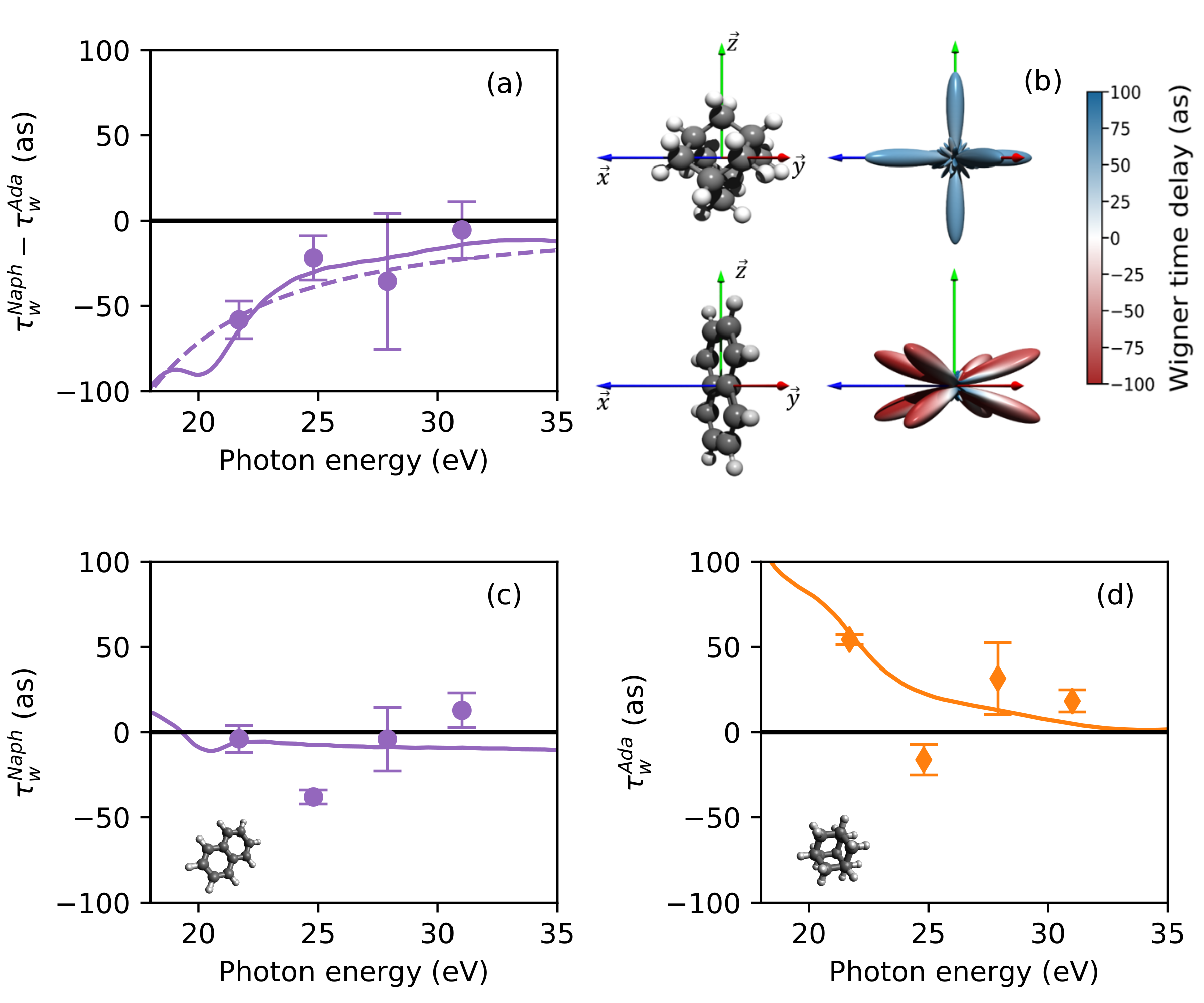}
\caption{{\bf Attosecond photoionization delays in naphthalene and adamantane.} {\bf (a)} Difference in Wigner delays between naphthalene ($\tau_w^{{\text {Naph}}}$) and adamantane ($\tau_w^{{\text {Ada}}}$). Experiments (purple dots), SE-DFT calculations (solid line) and model predictions (dashed line). {\bf (b)} Calculated ionization time delays in the molecular frame at a photon energy of 21.7~eV (the radial distance corresponds to the cross-section and the color to the value of the Wigner delay). 
Experimental absolute Wigner delays for {\bf (c)} naphthalene  (purple dots) and {\bf (d)} adamantane (orange triangles). The solid lines and the 3D plots show the results from the SE-DFT calculations for the HOMO orbital. The uncertainty for each measurement corresponds to 68~\% confidence interval, the statistical error is based on the fitting error. 
\label{Fig2}}
\end{figure*}

Following Wigner theory \cite{WignerIonizTime}, the phase induced by the XUV photoionization $\Delta \phi_{mol}$ can be interpreted as a delay ($\tau_w$) in the photoionization process via $\tau_w=\hbar~\frac{\Delta \phi_{mol}}{\Delta E_{HH}}$, where $\Delta E_{HH}$ is the energy difference between two adjacent harmonics. This is shown in Fig.~\ref{Fig2}(a), where a significant Wigner-delay-difference between naphthalene and adamantane, of the order of several tens of attoseconds, is observed over a $13$~eV photon energy range. We have compared these measured values with the results from the SE-DFT calculations (see Methods for details). The quantitative agreement between the experimental values and theory is excellent, which supports the validity of the measurements. Given that the two molecules have comparable sizes and composition, this difference in the time delays is quite remarkable. The absolute delay $\tau_w$ for the two molecules is obtained by using the calculated values of $\Delta \phi_{2h\nu}$ for argon from Ref. \cite{MauritssonPRA2005} and the $\tau_{cc}$ values from Ref. \cite{Dahlstrom2012}. The results are given in table~\ref{SummaryIonizTime} and shown in Fig.~\ref{Fig2}(c) and (d) for naphthalene and adamantane, respectively, together with the results of the SE-DFT calculations. Overall, the theoretical Wigner delays for the adamantane molecule stays positive or close to zero for the entire photon energy range whereas for naphthalene they are mostly negative. The trends of the experimental data are very well reproduced by the calculations. For the data close to 25~eV, the discrepancy might arise from Fano resonances excited by H17 in Ar \cite{NcommFano2016}, so that this discrepancy vanishes when the delay difference between two molecules is considered. To rule out any effect originating from the use of the NIR pulse as a dressing field, we have measured the $\tau_w$ values at various NIR intensities (see Fig.~\ref{Statistics_raw} of Methods). As can be seen, the measured Wigner delays do not depend on the intensity of the dressing field, showing the robustness of our experiment. In addition, the delay difference between naphthalene and adamantane is also negative for inner valence orbitals as shown in Fig.~\ref{Deconvol} of Methods.

The role of the 2D nature of the naphthalene molecule becomes apparent by looking at the Wigner delays obtained from the SE-DFT calculations in the molecular frames (see Fig.~\ref{Fig2}(b)). For adamantane, the $\tau_w$ extracted from the angle resolved ionization amplitudes in the molecular frame are found to be largely positive over all  photoemission directions. For naphthalene, they are mainly negative, especially along the direction for which the ionization probability is maximum. Because the molecules are randomly oriented in the experiment, there is no angular dependency in the measured time delays (see Methods). Nevertheless, averaging over the molecular orientation still results in negative time delays.

Considering photoionization as a half-scattering process, the interaction between the ejected photoelectron and the positively charged molecular cation can be described by an analytical model based on the first-order Born approximation. We note here that adamantane is an almost spherically symmetric molecule with its point group being $T_d$. Thus, one can assume that for this molecule the dominant interaction between the ejected electron and the residual molecular ion is nearly pure Coulombic, \textit{i.e.}, proportional to $1/r$, where $r$ is the distance between the emitted electron and the center of mass of the molecule. This is no longer true for a molecule with reduced symmetry, such as naphthalene (point group: $D_{2h}$). In this case, the interaction potential between the electron and the residual cation can have significant contributions from higher order terms in the multipole expansion:
    $V(r)=1/(4\pi\varepsilon_0)\left[e/r + C/r^2 + D/r^3+\dots \right ]$,
 where $C$ and $D$ are the dipole and quadrupole contributions, respectively. Because of the azimuthal symmetry of naphthalene, the dipole contributions can be neglected. The next dominant contributions is therefore that of the quadrupole term. 
 The Wigner delay associated with the quadrupole contribution in the direction perpendicular to the molecular plane can be written as (see Methods):
\begin{equation}
    \tau_{\textrm{q}}\approx - \frac{B_1}{\epsilon_{e}^{3/2}} -\frac{B_2}{S_n^2\epsilon_{e}^{5/2}}.
    \label{model}
\end{equation}
The terms $B_1$ and $B_2$ are estimated for naphthalene to be around $2$~fs$\cdot$eV$^{3/2}$ and $1.25$~fs$\cdot$eV$^{5/2}\cdot$nm$^4$, respectively. Here, $S_n$ is the area of the surface. Note that $\tau_q$ is always negative, indicating that the effect of the delocalization of the created hole is to decrease the photoionization delay by a quantity that depends on the spatial extension of the hole. In other words, measuring the time delay in extended 2D molecular systems gives access to the size of the hole created at the moment of ionization.  

By neglecting the higher order terms in the potential, the delay in the 3D molecule (adamantane) is governed by the Coulomb term, while the quadrupole term determines the delay in 2D systems. Therefore, the difference between the delay measured from the 2D and 3D molecules corresponds to the quadrupole contribution. Assuming that the hole created at the instant of ionization has the same surface area as that of the molecule, one could compare the experimental results with the model. In the case of naphthalene, $S_n=0.34$~nm$^2$, and we could obtain $\tau_q$ as a function of the electron kinetic energy. As shown in Fig.~\ref{Fig2}(a), $\tau_q $ is in very good agreement with the measured delays over an energy range of 13~eV. 

\begin{figure*}[!htb]
\includegraphics[width=\textwidth]{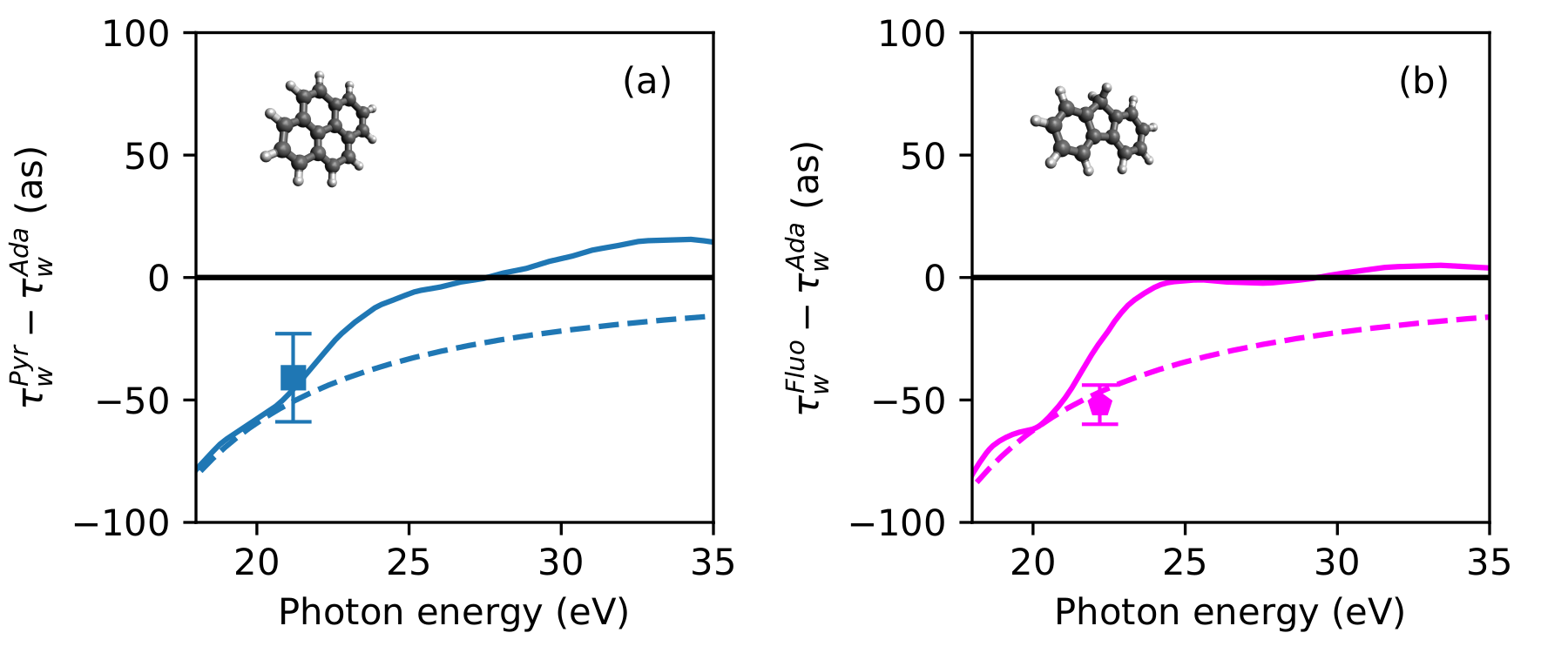}
\caption{{\bf Attosecond time delays in Fluorene and Pyrene.} 
Wigner time delay differences for {\bf (a)} pyrene and {\bf (b)} fluorene from experiment (symbols), SE-DFT calculations for the HOMO orbital (solid line) and the model (dashed line).
\label{Fig3}}
\end{figure*}

\begin{figure*}[!htb]
\includegraphics[width=\textwidth]{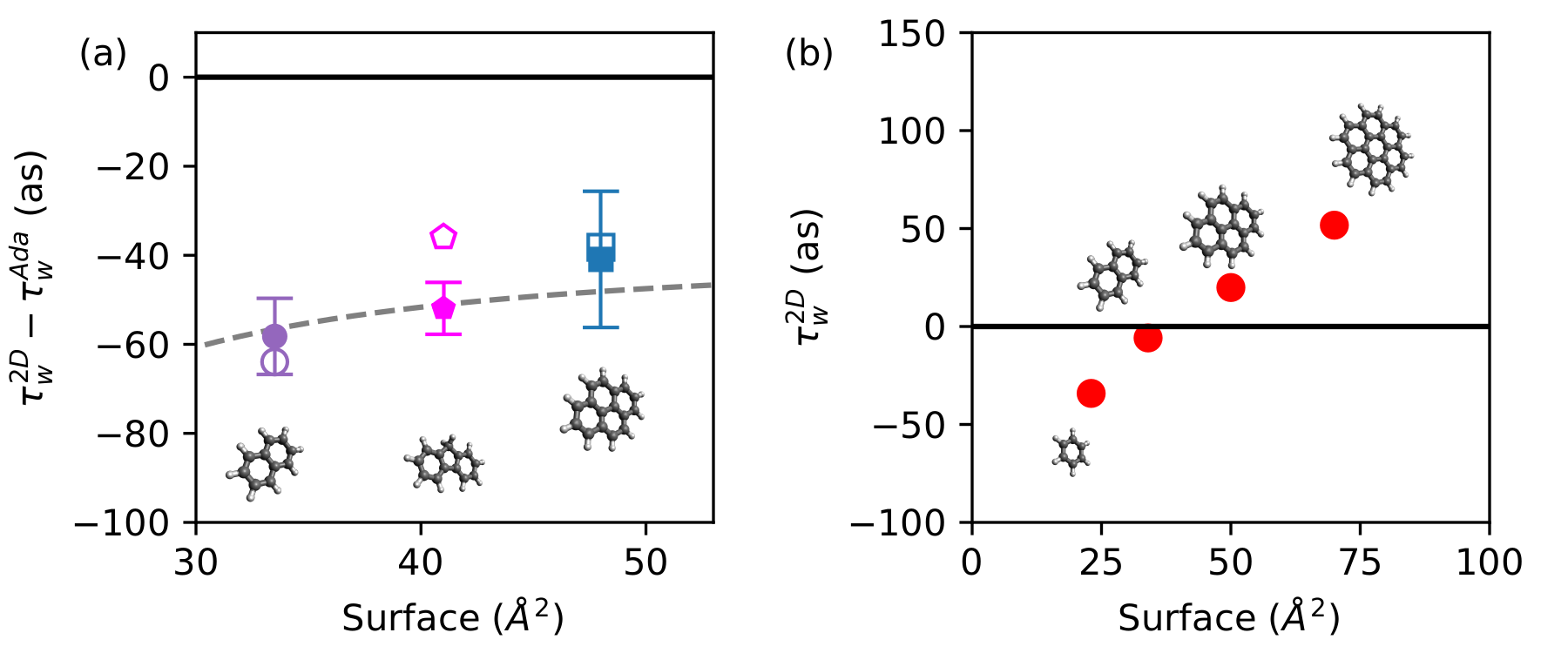}
\caption{{\bf Attosecond time delays in 2D molecules as a function of hole surface.} 
{\bf (a)} Difference of the ionization time delay for three 2D molecules of different size for a photon energy of 21.7~eV (referenced to adamantane). Experiments (symbols), SE-DFT calculations for the HOMO orbital (empty symbols) and model (dashed line).
{\bf (b)} Absolute Wigner delays for different PAHs as a function of the hole surface as predicted by the SE-DFT calculations for a photon energy of 21.7~eV for the HOMO orbitals.
\label{Fig4}}
\end{figure*}

To investigate the generality of our findings, the same experimental procedure has been applied to two larger 2D molecules: pyrene and fluorene. Again the measurements are in good agreement with the SE-DFT calculations and with the model (see Fig.~\ref{Fig3}(a) and Fig.~\ref{Fig3}(b)), which confirms the conclusions drawn from the analysis of the naphthalene data. The measured delay differences for the three 2D molecules are presented as a function of the hole size in Fig.~\ref{Fig4}(a), in excellent agreement with the model. As expected the delay increases with the 2D hole size and matches the delay predicted from the surface of the molecule. This behavior is also illustrated in a series of calculations performed on other 2D molecules (benzene and polycyclic aromatic hydrocarbons (PAH)) presented in Fig.~\ref{Fig4}(b). It shows that the total delay varies from -40 as to + 50~as when the hole surface increases from 25 to 75 \AA{}, highlighting the connection between the size and symmetry of the spatial extension of the created hole in even larger 2D systems. To compare $\tau_q$ and $\tau_{w}$, we have explicitly used the size of the hole $S_n$ created in 2D molecules. Alternatively, one could also use Eq.~\ref{model} to obtain the size of the hole in other types of 2D structures from the measured time delays. 

Our results show that beyond resonance effects investigated in small quantum objects, ionization delays in extended systems are not simply governed by the molecular size but symmetry plays also a major role, even in a flat non-resonant electronic continuum. Recent experiments showed the importance of `correlation bands' in femtosecond non-adiabatic dynamics of holes created upon photoionization of large 2D molecules \cite{HerveNPhys2021}. Here we demonstrate that attosecond metrology can serve as a probe of the spatial properties of the hole, with angstrom accuracy, at the instant of ionization. 
Despite the complexity of the targets studied here, we could extract the photo-ionization time delays associated with several molecular orbitals and derive an analytical trend based on the general multipole expansion of the interaction potential between the residual ion and the emitted photoelectron.
As a perspective, the pre-alignment of the molecular sample coupled with the angle-resolved ionization time could provide channel-selectivity and 3D reconstruction of the hole in attosecond time-resolved molecular photoionization.
Also, given the rapid developments in attosecond metrology in condensed phase \cite{Tao2016,Siek2017}, our findings could inspire future experiments for molecules deposited on a surface making the connection between microscopy technology with attosecond precision \cite{Garg2022}.

\vspace{1cm}
{\bf Acknowledgments\\}
We thank M. Herv\'{e}, E. Constant, G. Karras for fruitful discussions. All calculations were performed at the Centro de Computaci\'{o}n Cient\'{\i}fica (CCC) of the Universidad Aut\'{o}noma de Madrid, and the MareNostrum supercomputer of the Red Espa\~{n}ola de Supercomputaci\'on. {\bf Funding:} We acknowledge the support of CNRS, ANR-16-CE30-0012 ``Circ\'e'', ANR-15-CE30-0001 ``CIMBAAD'', the F\'ed\'eration de recherche Andr\'e Marie Amp\`ere and the European COST Action AttoChem (CA18222). CMGC, EP, AP and FM are supported by the  Ministerio de Ciencia e Innovaci\'{o}n (MICINN) project PID2019-105458RB-I00 and the Comunidad de Madrid project FULMATEN (Ref. Y2018\/NMT-5028).  FM acknowledges support from the `Severo Ochoa' Programme for Centres of Excellence in R\&D (CEX2020-001039-S) and the ``Maria de Maeztu'' Programme for Units of Excellence in R\&D (CEX2018-000805-M). \\

{\bf Author contributions:} V.L. and F.L. conceived the project. A.B., A.M., C.L.G., Y.H., S.N., F.L., and V.L. performed the experiment. V.L. analyzed the data. S.N. developed the analytical model. C.M.G.C, M.L-.A. and E.P. performed the theoretical calculations under the supervision of A.P., P.D., and F.M. The manuscript is written by V.L., S.N., and F.L. with inputs from all the authors. V.L., F.M., and F.L. supervised the project.\\ 

{\bf Competing interests}: The authors declare no competing interests.\\

\bibliographystyle{nature.bst}
\nocite{TitlesOn}

\newpage
\section*{Methods}
\begin{figure*}[!htb]
\includegraphics[width=\textwidth]{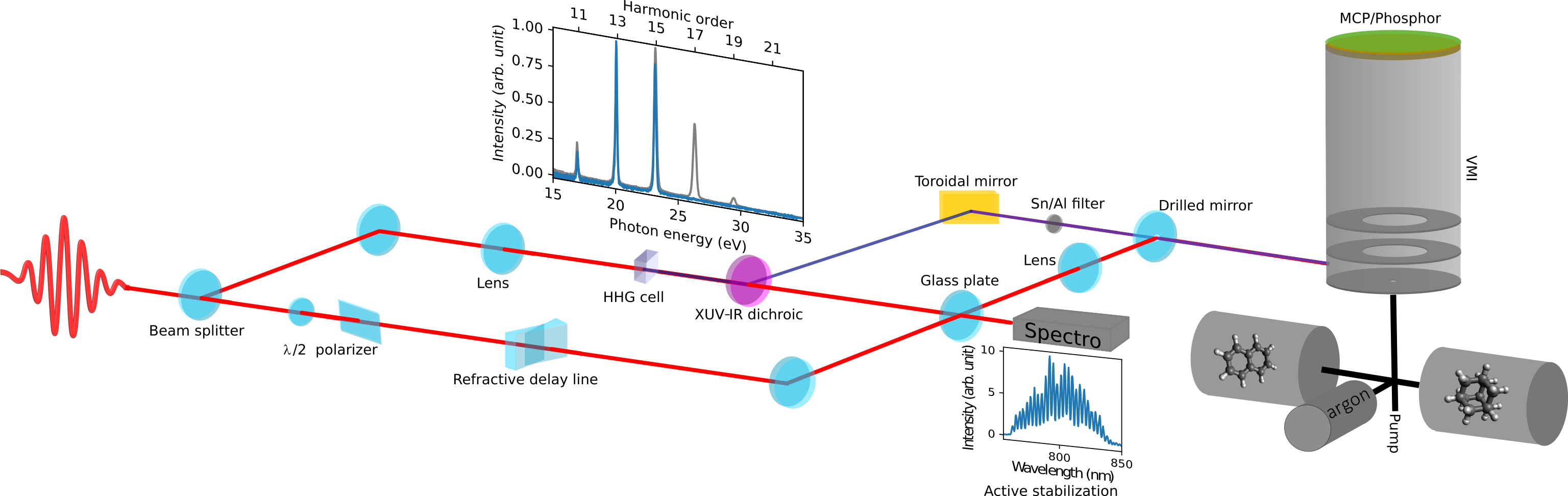}
\caption{\textbf{Experimental setup}. The two main components of the setup: a Mach-Zehnder interferometer and a velocity map imaging spectrometer (VMIS). See text for details.
\label{SetupAtto}}
\end{figure*}

{\bf Experimental details\\}
The experimental setup is shown in Fig.~\ref{SetupAtto}. A $2$~mJ, $25$~fs near infra-red (NIR) pulse at $5$~kHz repetition rate having central wavelength around $800$~nm from a commercially available system (Legend-Elite Duo, Coherent) is sent into a Mach-Zehnder interferometer. In the `pump' arm, it is up-converted into extreme ultraviolet (XUV) radiation via high-order harmonic generation (HHG) using xenon or krypton. The driving beam is filtered at first via a dichroic mirror, followed by a metallic filter. The resulting XUV spectra containing either three odd-order harmonics: $11-15$ for a $200$~nm Sn filter (top inset in light blue), or higher odd-order harmonics distribution as a function of the generation gas and the phase matching conditions filtered by a $200$~nm Al foil (top inset in gray for the study of SB18) are focused by a toroidal mirror to the interaction region of a velocity map imaging spectrometer (VMIS) \cite{eppink1997RSI}. In the `probe' arm, the energy of the NIR pulse at $800$~nm is tuned by a half-waveplate polarizer assembly. It is delayed in time from the XUV-pulse by a refractive delay-line composed of two small-angle prisms that do not affect significantly the duration of the NIR pulse. The NIR pulse is then focused by a lens with $1$-m focal length to the same point as the XUV pulse. The spatial recombination between the two pulses is realized by a $3$~mm drilled mirror. To prevent any temporal drift, the interferometer is actively stabilized in a close loop. The targets (adamantane, naphthalene, argon or, a mixture of one of the molecules and argon) are introduced as an effusive jet through a $200~\mu$m hole at the center of the repeller electrode in the VMIS. 

\begin{figure*}[!htb]
\includegraphics[width=\textwidth]{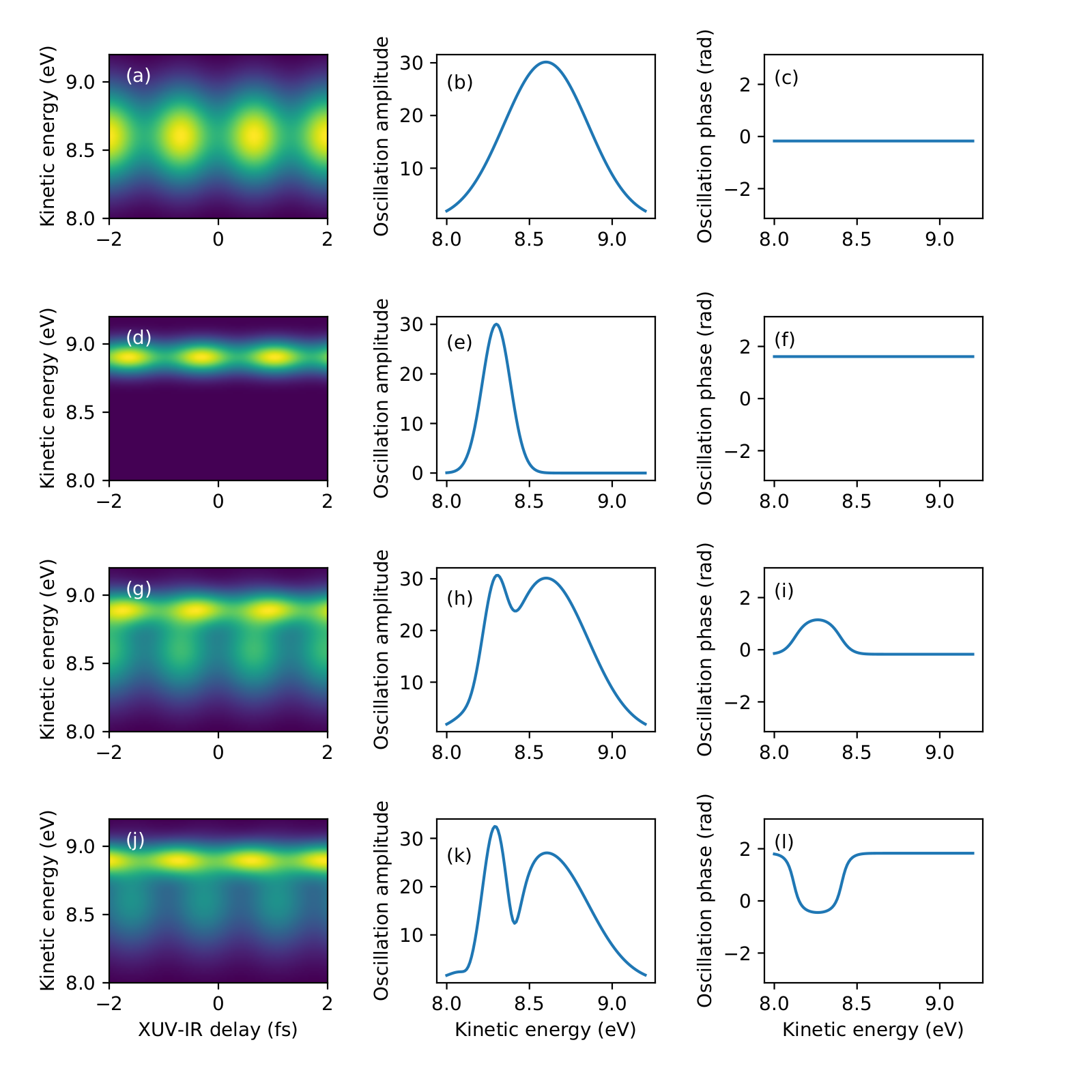}
\caption{\textbf{Relative temporal calibration using gas mixture}. {\bf (a)} and {\bf (d)} represent the simulated RABBIT-map of two individual targets. The direct sum of the simulations {\bf (a)} and {\bf (d)} is shown in {\bf (g)}. The RABBIT-map of the gas-mixture, considering a random temporal drift between each measurements and a different partial pressure, is shown in {\bf (j)}. The corresponding 2$\omega_0$ oscillation amplitudes {\bf (b,e,h,k)} and phases {\bf (c,f,i,l)} of the measurements {\bf (a,d,g,j)} are shown in the central and the right column, respectively.
\label{RescaleTime}}
\end{figure*}

\begin{figure*}[!htb]
\centering
\includegraphics[width=\textwidth]{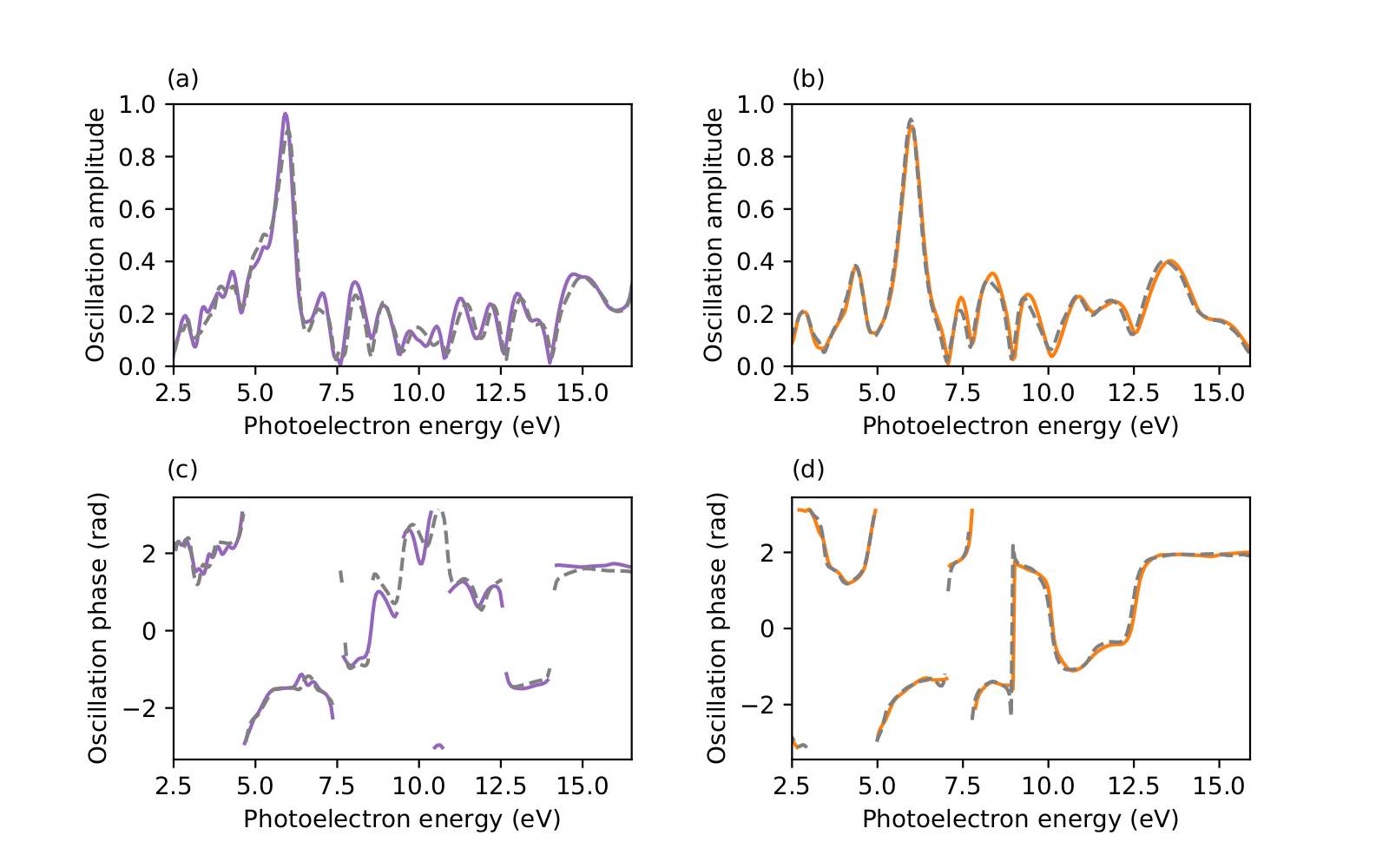}
\caption{\textbf{Temporal rescaling of the RABBIT measurements for naphthalene and adamantane}. Oscillation {\bf(a-b)} amplitudes and {\bf(c-d)} phases at twice the fundamental frequency (2$\omega_0$) for {\bf(a,c)} naphthalene and  {\bf(b,d)} adamantane. The solid lines are the measurements of molecule and argon together in a mixture and the dashed lines are the corresponding reconstructions (Eq.~\ref{ComplexOscillSum}) of the oscillation from measurements in pure molecular and atomic sample taken separately.
\label{CompareOscillAdaNaph}}
\end{figure*}

\vspace{1cm}
{\bf Relative temporal calibration\\}
Despite an active stabilization of the delay-line in the main part of the interferometer, there can be a long-term temporal drift. This may affect the individual measurements when the sample is being changed. Because of this temporal drift and a slightly different target pressure in the interaction region, the direct sum of the two individual measurements, one for a molecule and one for argon, does not match exactly the measurement having both of them together. This is shown in Fig.~\ref{RescaleTime}, where RABBIT maps are simulated for two different gas targets with a partial overlap of their contributions. By numerically shifting the individual measurements in time and weighting with their oscillations amplitudes, it is possible to unambiguously reconstruct the oscillation observed in the gas mixture. This is performed by extracting the complex amplitude and phase of the oscillation $\tilde{R}_{2\omega}$ of the individual measurements ($A$ and $B$) as a function of the kinetic energy $E$ ($\tilde{R}_{2\omega,A}(E)$ and $\tilde{R}_{2\omega,B}(E)$), and multiplying them by a complex factor ($\tilde{Z}_{A}$ and $\tilde{Z}_{B}$ respectively) to reproduce the oscillation of the mixture ($\tilde{R}_{\textrm{A+B}}(E)$):
\begin{equation}\tag{S1}
\tilde{R}_{2\omega,\textrm{A+B}}(E)=\tilde{Z}_{\textrm{A}} \tilde{R}_{2\omega,\textrm{A}}(E) + \tilde{Z}_{\textrm{B}} \tilde{R}_{2\omega,\textrm{B}}(E).\label{ComplexOscillSum}
\end{equation}
The factor $\tilde{Z}=|Z|e^{i\phi_Z}$ involves both the oscillation amplitude and phase of the individual oscillations. Here, $\phi_Z$ corresponds to a shift in the temporal domain ($\phi_Z=2\omega_0 t_\textrm{shift}$). The $\tilde{Z}$ coefficients are optimized with a least-squared algorithm in their Cartesian representation. 
We used this method to temporally re-scale the individual measurements and extract their relative phases free of temporal jitter. 
The optimizations with time re-scaling for the experimental data are shown in Fig.~\ref{CompareOscillAdaNaph} for naphthalene-argon mixture (Naph-Ar) and adamantane-argon mixture (Ada-Ar), in both cases using the Sn-filter. 
Following this method, all the measured oscillations contribute to the calibration. 

\begin{figure*}[!htb]
\centering
\includegraphics[width=\textwidth]{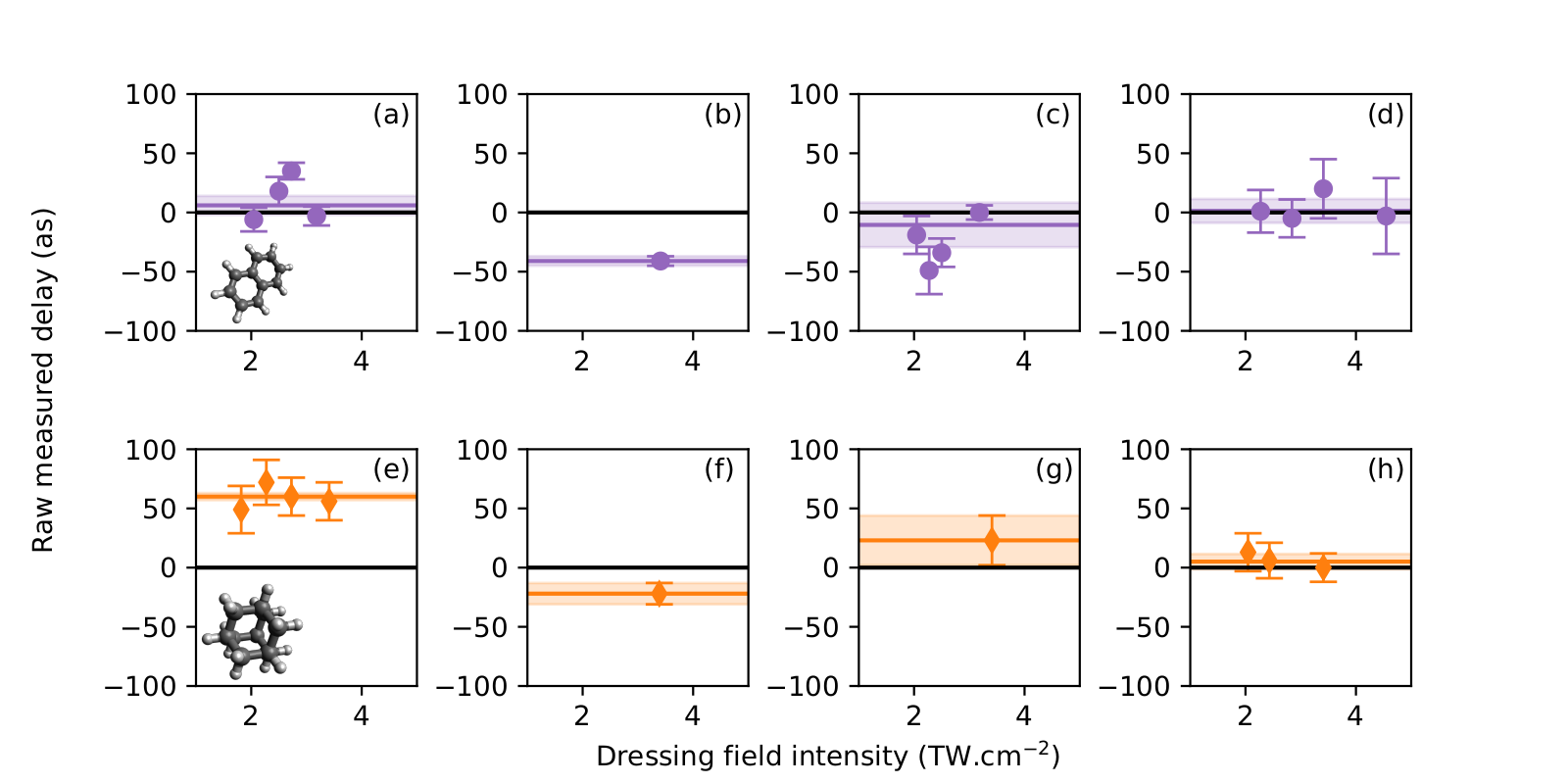}
\caption{\textbf{Experimental two-photon delays}. The experimental values for {\bf(a-d)} naphthalene and {\bf(e-h)} adamantane for sidebands SB14, SB16, SB18 and SB20 respectively.
\label{Statistics_raw}}
\end{figure*}

\begin{figure*}[!htb]
\centering
\includegraphics[width=0.7\textwidth]{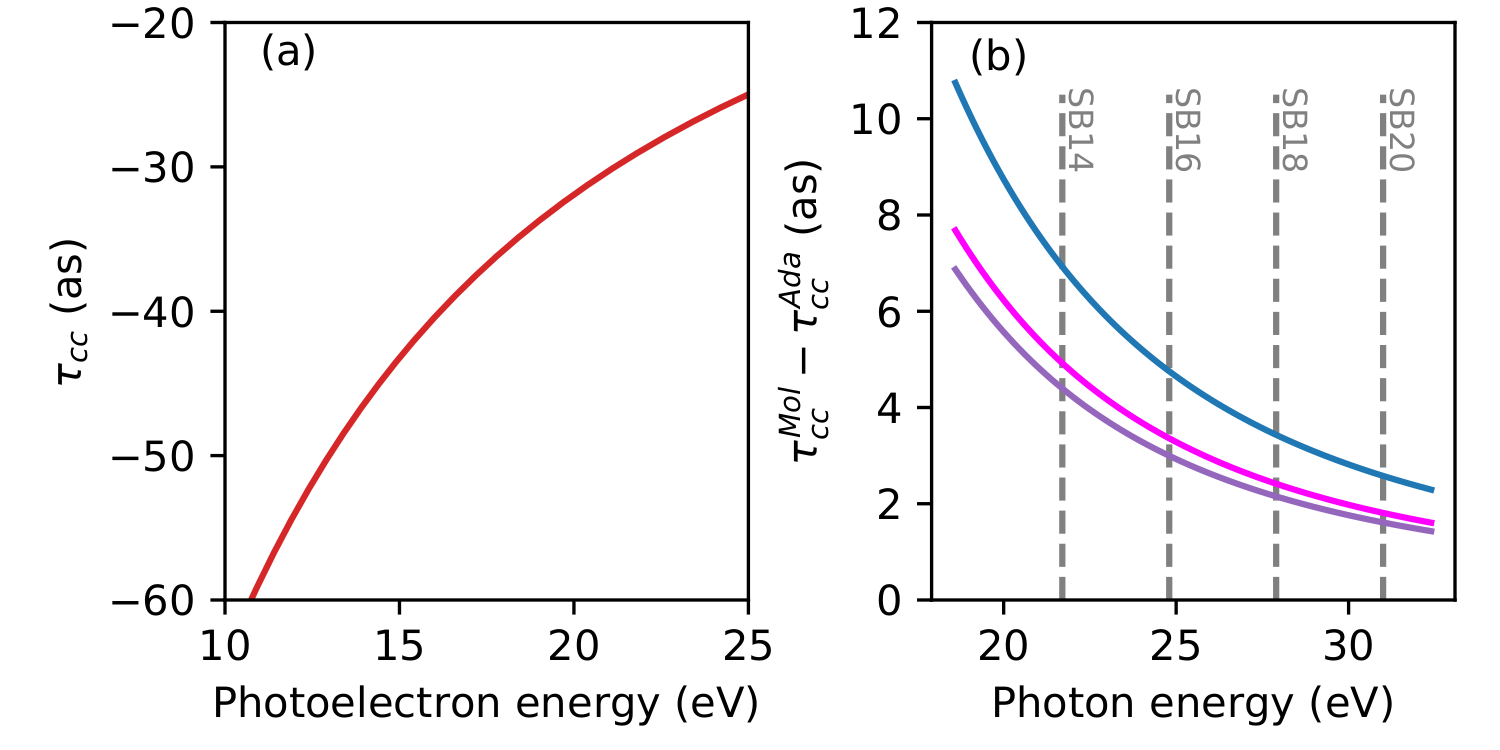}
\caption{\textbf{Continuum-continuum contribution}. {\bf(a)} Theoretical continuum-continuum delay $\tau_{cc}$ from Ref. \cite{Dahlstrom2012SM} as a function of the photoelectron energy (solid line). {\bf(b)} $\tau_{cc}$ correction for  naphthalene (purple),  fluorene (fuchsia) and  pyrene (fuchsia) calculated from (a) considering the differences in the ionization potentials compared to adamantane.
\label{phicc}}
\end{figure*}

\vspace{2cm}
{\bf Determining the absolute Wigner delay \\}
The experimental two-photon delays for both molecules are shown in Fig.~\ref{Statistics_raw}. We used the measurements in argon as a calibration to compare the oscillation phases between adamantane and naphthalene under the same experimental conditions. 
Because the ionization potential of adamantane: $9.3$~eV is similar to that of naphthalene: $8.15$~eV \cite{NIST2001SM}, the respective contributions of $\Delta\phi_{cc}$ are very close to each other \cite{Dahlstrom2012SM}, within $5$~attoseconds for the sideband $14$  (see Fig.~\ref{phicc}). To extract the absolute ionization time delay, the calculated phase for two photon transition ($\Delta\phi_{2h\nu}$) in argon \cite{MauritssonPRA2005SM} were used as a reference. The experimentally measured phase $\phi$ can be decomposed as:
\begin{equation}\tag{S2}
    \phi=\left(\underbrace{\Delta\phi_{2h\nu}^{mol}}_{\Delta\phi_{mol}+\Delta\phi_{cc}}+\cancel{\Delta\phi_{HH}}\right)-\left(\Delta\phi_{2h\nu}^{Ar}+\cancel{\Delta\phi_{HH}}\right),
\end{equation}
where the attochirp $\Delta\phi_{HH}$ gets canceled between the two targets in a gas mixture. The absolute molecular Wigner time delay ($\Delta\phi_{mol}$) can therefore be obtained using the following equation
\begin{equation}\tag{S3}
    \Delta\phi_{mol}= \phi-\Delta\phi_{cc}+\Delta\phi_{2h\nu}^{Ar}.
\end{equation}
The harmonics 13 and 15 are almost isolated in the spectrum (see inset in Fig.~\ref{SetupAtto}). Hence, the phase of the sideband 14 is used as a reference in the measurements. All the following measurements have the same $\Delta \phi_{HH}$ that cancels the phase difference between oscillations. 
The observed phase differences for the oscillation at the highest kinetic energy of the RABBIT signal between adamantane and naphthalene can only originate from a difference of molecular phase ($\Delta\phi_{mol}$), induced by ionization of their respective HOMO states by the harmonics 13 and 15. As shown in the main text, the phase-difference $\Delta\phi$ can be converted in time following Wigner \cite{WignerIonizTimeSM} using $\tau=\hbar\Delta\phi/\Delta E_{HH}$.
The different quantities necessary to reconstruct the absolute ionization time delays are summarized in Table~\ref{SummaryIonizTime}.

\begin{figure*}[!htb]
\centering
\includegraphics[width=\textwidth]{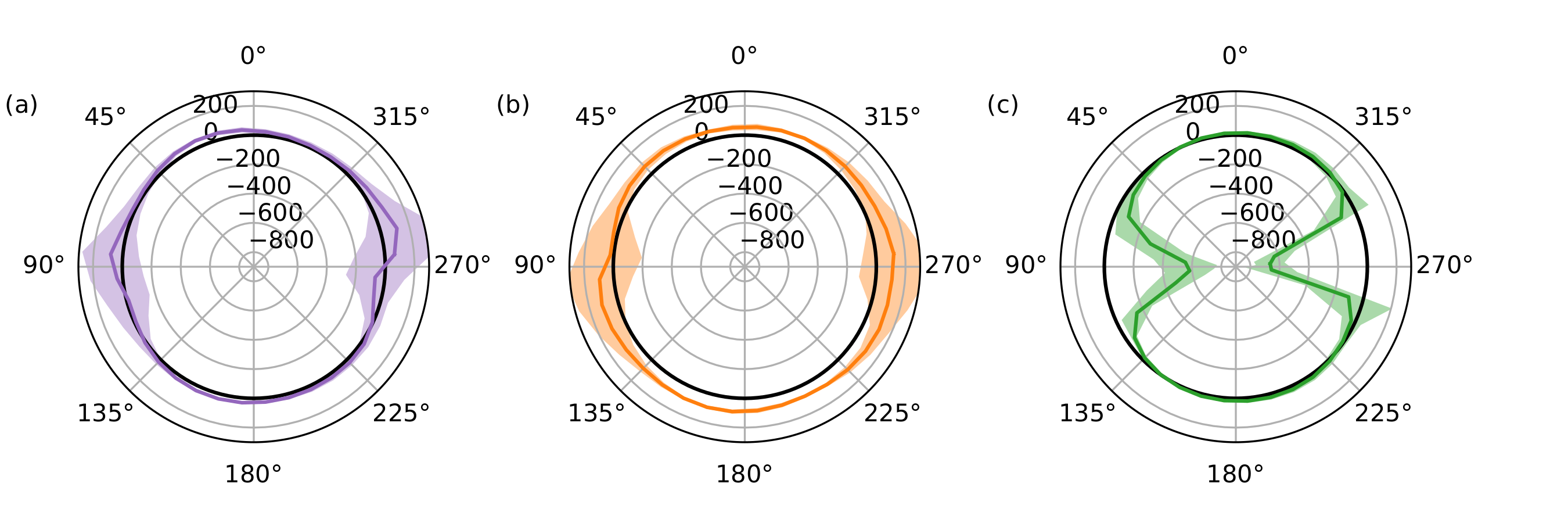}
\caption{\textbf{Angle-resolved delays.} Polar representation of experimental two-photon delays in attoseconds for {\bf(a)} naphthalene, {\bf(b)} adamantane and {\bf(c)} argon in case of sideband 14.
\label{Angularly_Resolved_Time_delay_paper}}
\end{figure*}

\vspace{1cm}
{\bf Angle-resolved photoionization time delay\\}
The ionization time delay has also been extracted as a function of the emission angle with respect to the polarization axis. The Abel-inverted images were sampled in steps of 10$^\circ$ providing $36$ independent RABBIT spectrograms. We performed Fourier transform on each of the measurements to extract the phase of the oscillation at 2$\omega_0$. These oscillations are calibrated on the SB14 of argon through the relative temporal calibration presented above. The result is displayed in Fig.~\ref{Angularly_Resolved_Time_delay_paper}, where the analysis is performed on one specific measurement (the spectrogram shown in Fig.~\ref{Fig1} of the main text). The error-bars correspond to the phase dispersion within the width of the photoelectron band. It can be seen that the photoionization time delay is almost independent of the emission angle for both naphthalene and adamantane molecules. For argon, the angular dependency is in agreement with previously reported measurements \cite{Cirelli2018} where the ionization time drops sharply perpendicular to the polarization axis.

\begin{figure*}[!htb]
\centering
\includegraphics[width=\textwidth]{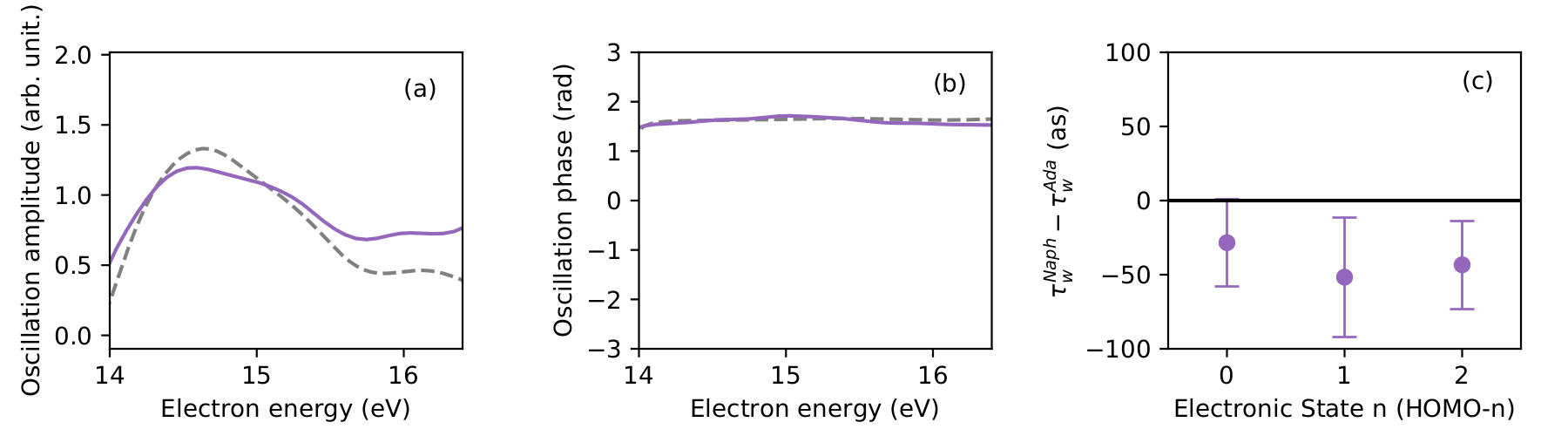}
\caption{\textbf{Example of reconstruction of the oscillations for highest kinetic energy photoelectrons.} Oscillation {\bf(a)} amplitude and {\bf(b)} phase at twice the fundamental frequency (2$\omega_0$) located at the position of SB16 in naphthalene reflecting the oscillation of SB14 with a shift of $\pi$. The solid line is the actual measurement in naphthalene and the dashed line is the sum of all three contributions shown in {\bf (c)}.
\label{Deconvol}}
\end{figure*}

\vspace{2cm}
{\bf Deconvolution of the states\\}
Naphthalene has three well separated outer valence orbitals with binding energies 8.15~eV (HOMO), 8.9~eV (HOMO-1) and 10.7~eV (HOMO-2), as shown in Ref. \cite{JACS72}. These states are within 3~eV of each others and have similar photoionization cross-sections in the photon energy range considered here. The highest oscillations measured in the RABBIT map correspond to that of the main-band (MB) 15 and its one-photon replica at the position of SB16. Hence, each electronic state may produce photoelectron oscillations at MB15 and SB16 that oscillate in phase with a constant ratio. Because the electron kinetic energy is different for the three states, the spectral congestion is low allowing the deconvolution of the oscillations \cite{Jordan2018}. The variation of oscillation phase as a function of the kinetic energy within the width of a particular band can be extracted from the measurements in argon. According to the measurements in argon, the oscillation phase is flat within the width of the peak.
The RABBIT oscillations at the highest kinetic energy range can be adjusted by the above described contributions where each electronic state has a constant phase for MB15 and SB16. The global ratio between MB15 and SB16 is fixed for the three contributions but the global magnitude of each contribution is used as a free parameter. This method allowed us to extract the phase for the three outer states independently (see Fig.~\ref{Deconvol}). 

\vspace{1cm}
{\bf \textit{ab initio} calculations\\}
We have used the B-spline static exchange density functional theory (SE-DFT) method \cite{Stener1999,Toffoli2002} that has been successfully applied to describe photoionization of other molecules of similar size \cite{ReviewNisoli2017}. In brief, we have evaluated dipole transition matrix elements for the $\alpha$ ionization channel in the length gauge
\begin{equation}\tag{S6}
    d^{\alpha}_{q,\ell,\mu} = \langle\varphi^-_{q,\ell,\epsilon_e}(R;\mathbf{r}_{q_j})|r_jY_1^\mu|\varphi_\alpha(R;\mathbf{r}_{q_j})\rangle,
\end{equation}
where $\varphi_\alpha$ is the Kohn-Sham (KS) orbital from which the electron is ejected into the continuum,
$\varphi^-_{q,\ell,\epsilon_e}$ is the final one-electron continuum wave function with the proper incoming asymptotic conditions for a photoelectron with energy $\epsilon_e$, angular momentum $\ell$, and orbital symmetry $q$, $\mathbf{r}_{q_j}$ is the coordinate of electron $j$, $R$ represents the nuclear coordinates at the equilibrium position, $Y^\mu_1$ is the spherical harmonic with $\ell=1$ and $\mu$ denotes the spherical component of the dipole operator. One-photon ionization cross sections are simply given by the square of this matrix element with the appropriate kinematic factors. We have evaluated the corresponding Wigner delays from the formula \cite{WornerIoniztime}

\begin{align}
    \tau_1^\alpha(R,\epsilon_e)&= \frac{1}{8\pi^2}\int d\Omega_M\int d\Omega_e \frac{\sigma_\alpha(R,\epsilon_e,\Omega_M,\Omega_e)}{\sigma_\alpha(R,\epsilon_e)}\nonumber\\ 
    & \times \frac{\partial}{\partial\epsilon_e}\textrm{arg}\left( \sum_{q,\ell,\mu}(-i)^\ell e^{i\eta_\ell(\epsilon)}D^1_{\mu0}(\Omega_M)d^\alpha_{q,\ell,\mu}(R,\epsilon_e)\sum_m b^q_{\ell,m}Y^m_\ell (\Omega_e) \right), \tag{S7}
    \label{DFTwigner}
\end{align}
where $\Omega_M$ and $\Omega_e$ represent, respectively, the set of Euler angles of the polarization vector and the emitted electron relative to the molecular frame,  $\eta(\epsilon)$ is the Coulomb phase, $D_ {\mu 0}^1$ the Wigner rotation matrix for linearly polarized light, $b_{\ell,m}^q$ the coefficients of the angular expansion in real spherical harmonics $Y_\ell^m(\Omega_e)$, $\sigma_\alpha(R,\epsilon_e,\Omega_M,\Omega_e)$ and $\sigma_\alpha(R,\epsilon_e)$ are the angularly resolved and total photoionization cross sections, respectively.
Equation \ref{DFTwigner} is the average over electron emission angle and molecular orientation of the angle-dependent photoionization Wigner delay \cite{WignerIonizTimeSM, SmithIonizTime, RMP2015SM}.

The KS orbitals were determined by diagonalizing the field-free KS Hamiltonian in a multicenter basis set of B-splines functions and symmetry-adapted real spherical harmonics \cite{Stener2007}. The KS potential was built with the LB94 exchange-correlation functional \cite{Leeuwen1994} and the electronic density was calculated by using the Amsterdam Density Functional package with a DZP basis set. The continuum wave functions were evaluated through an inverse iterative procedure in the former KS basis for each photoelectron energy, thus ensuring the proper incoming asymptotic behavior. The multicenter basis set consists of two types of expansion centers: one located at the center of mass of the molecule, with a large radius ($R^0_\textrm{max}=$75~a.u.) and a large number of angular momenta ($\ell^0_\textrm{max}$=26 for all molecules except for coronene and tetracene, where we have used $\ell^0_\textrm{max}=30$ and $32$, respectively), in order to describe accurately the long-range oscillatory behaviour of the continuum states, and additional expansions centered on the carbon atoms ($R^1_\textrm{max}=1.2$~a.u. and $\ell^1_\textrm{max}=1$) to improve the description of the cusp of the wave function in the vicinity of the nuclei. For all systems, in the one-center expansions, we have used 250 B-spline functions and in the off-center expansions, 5 B-spline functions. 

\begin{figure*}[!htb]
\centering
\includegraphics[width=\textwidth]{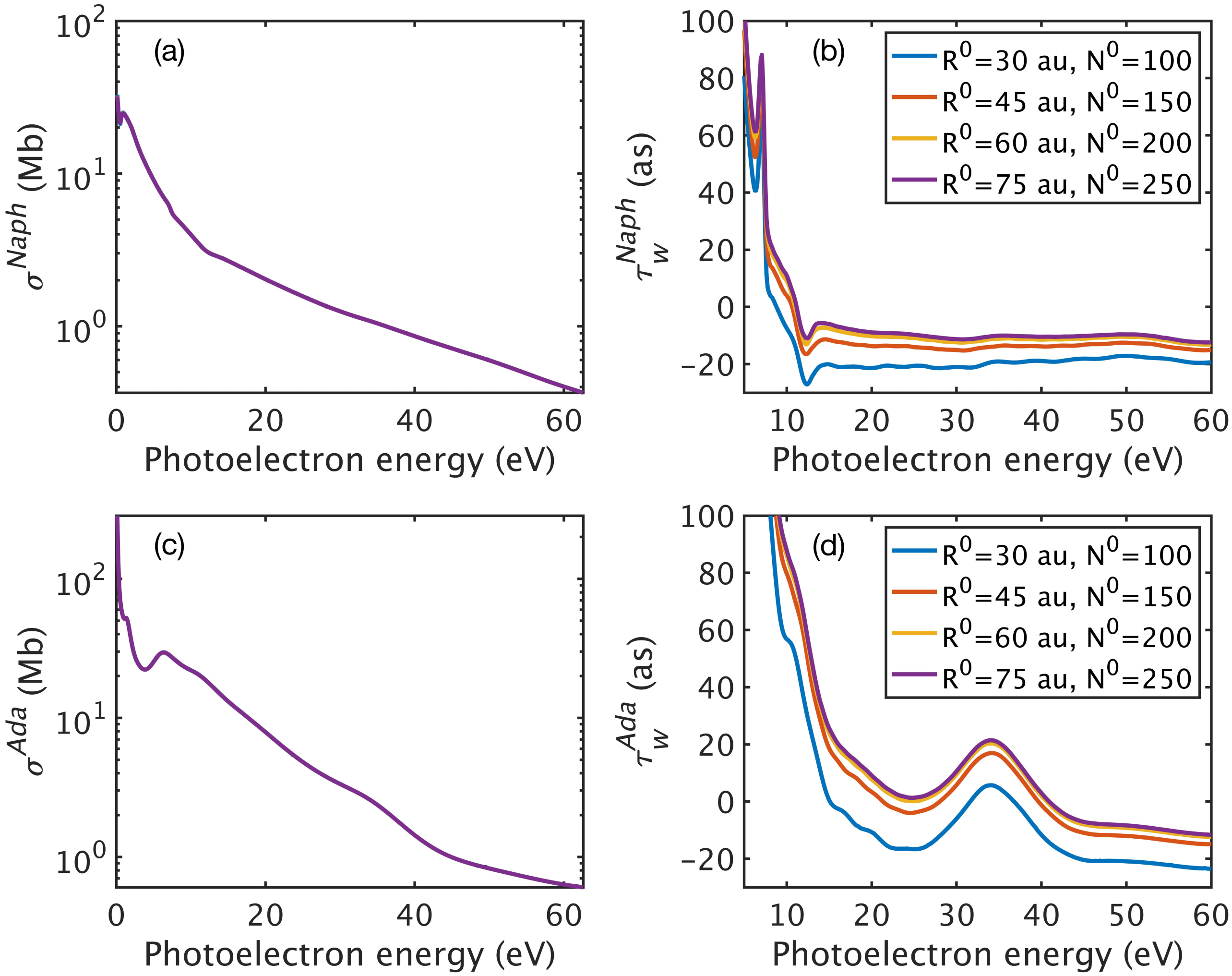}
\caption{\textbf{SE-DFT calculation for as a function of the basis size}. {\bf(a,c)} photoabsorption cross-section and {\bf(b,d)} their corresponding Wigner time delay for {\bf(a-b)} naphthalene and {\bf(c-d)} adamantane, respectively as a function of the basis size.
\label{ConvergenceSEDFT}}
\end{figure*}

The convergence of the calculated cross sections and Wigner delays with the box size and the number of B-spline functions is illustrated in Fig.~\ref{ConvergenceSEDFT} for the case of the naphtalene and the adamantane molecules. As can be seen, the calculated cross sections are nearly identical for all box sizes and numbers of B-spline functions indicated in the figures, while Wigner delays are more sensitive to these parameters and are only converged for the larger boxes (75 a.u.) and a larger number of B-spline functions (250). 

\vspace{1cm}
{\bf Analytical model\\}
For single photoionization of atoms and/or molecules, the interaction between the resulting positive ion and the emitted photoelectron is usually considered as an attractive Coulomb potential. It follows from the spherical symmetry attributed to the residual ion, effectively reducing it to a point charge. Since the Coulombic interaction between two point charges depends solely on their mutual distance, it cannot provide any information specific to the geometry or structure of the target system.

For an extended molecule with dimensions much larger than the Bohr radius, this is no longer a reasonable approximation, since a non-negligible contribution from the higher order multi-poles can appear. For polycyclic aromatic hydrocarbon (PAH) molecules with planar symmetry as those considered in this work, the system can be approximated as a 2D surface. Such a model has already been used to describe collision-induced electron transfer processes from PAH molecules to highly charged heavy ions \cite{Forsberg2013}. 

Let us consider that the created hole (charge $=+e$) is uniformly smeared over surface of the 2D system.  
In that approximation, solving Laplace's equation in the spherical polar co-ordinate, the electric potential at an arbitrary point $\left(r,\theta\right)$ can be written as, 
\begin{equation}\tag{S8}
    V_\textrm{surf}=\frac{e}{4\pi\varepsilon_0}\left[\frac{1}{r}-\frac{R^2}{4r^3}P_2(\cos\theta)+\dots\right],
\end{equation}
where $r>R$.
Perpendicular to the surface, we have,
\begin{equation}\tag{S9}
    V_\textrm{surf}\left(r,\theta=0\right)=\frac{e}{4\pi\varepsilon_0}\left[\frac{1}{r}-\frac{R^2}{4r^3}+\dots\right], \label{approxperp}
\end{equation}
The term additional to the Coulombic contribution ($1/r$) arises from quadrupole contributions due to the extended nature of the hole. The corresponding quadrupole moment, $\mathcal{D}$, is given by, 
\begin{equation}\tag{S10}
    \mathcal{D}= - \frac{e R^2}{4}
\end{equation}

Combining all of these, the potential energy of the emitted photoelectron moving in this potential can be written as:
\begin{equation}\tag{S11}
    V=-\frac{e^2}{4\pi\varepsilon_0}\left[\frac{1}{r}-\frac{R^2}{4r^3}+\dots\right].
\end{equation}
Note that in addition to the usual attractive Coulomb term, there is a short-range repulsive term in the potential energy due to the quadrupole contribution: $V_q=\gamma/r^3$, with $\gamma=e^2R^2/16\pi\varepsilon_0=\mathcal{D}^2/\varepsilon_0 S_n$. Here, $S_n$ denotes the area of the molecule.
Following Landau and Lifshitz \cite{landauBook}, for slow photoelectrons being dispersed in $V_q$, we can apply the first-order Born approximation to get the scattering phase shifts. For a spherically symmetric potential $V_q(r) = \gamma/r^3$, the  phase shift for each partial wave is described analytically by :
\begin{align}
    \tan\eta_\ell&= -\gamma\frac{\sqrt{\pi}  m_e}{2\hbar^2}\frac{1}{\Gamma\left(\frac{3}{2}\right)}\frac{\Gamma\left(\ell\right)}{\Gamma\left(\ell+2\right)}k, \nonumber \\ 
    \eta_\ell&= \tan^{-1}\left(-\gamma\frac{ m_e}{\hbar^2}\frac{k}{\ell(\ell+1)}\right), \tag{S12}
\end{align}
with $k$ being the wave-vector of the photoelectron, $m_e$ the mass of the electron, and $\ell$ ($>0$) the corresponding angular momentum. 
The corresponding short range (Wigner) delay can be obtained as, 
\begin{align}
    \tau_\ell&= 2\hbar\frac{d\eta_\ell}{dE}=2\hbar\frac{dk}{dE}~\frac{d\eta_\ell}{dk} \nonumber \\
          &= 2\frac{m_e}{\hbar k}\times \left( -\frac{\gamma~                m_e\hbar^2~\ell(\ell+1)}{\left(\hbar^2~\ell(\ell+1)\right)^2 + \gamma^2m_e^2k^2}\right) \nonumber \\
    &= -\frac{2~\gamma~m_e^2~\hbar~\ell(\ell+1)}{k\left(\hbar^2~\ell(\ell+1)\right)^2 + \gamma^2m_e^2k^3}.     \tag{S13}
\end{align}
The cross-sectional area associated with a given $\ell$-wave can be calculated semi-classically as follows: for an electron with momentum $p$ and impact parameter $b$, the classical angular momentum is defined as $L = pb$. Quantum mechanically it is defined as: $L = \hbar\ell$. Let us consider that the quantized angular momentum $\ell$ corresponds approximately to the impact parameter $b=L/p$ for $\hbar\ell < L <\hbar(\ell+1)$ i.e., for $\ell/k \leq b \leq (\ell+1)/k$.
Assuming the electron gets scattered with 100\% probability when entering in this annular region, the maximum cross-sectional area can be written as:
\begin{equation}
    \pi\left(\frac{\ell+1}{k}\right)^2-\pi\left(\frac{\ell}{k}\right)^2=\frac{(2\ell+1)\pi}{k^2}. \tag{S14}
\end{equation}
Using the approach used by Nussenzveig \cite{Nussenzveig72}, the Wigner delay averaged over all partial waves $\ell$ can be obtained as, 
\begin{align}
    \tau_q&= \frac{1}{\sigma}\sum_\ell \tau_\ell \frac{(2\ell+1)\pi}{k^2}, \nonumber  \\
    \tau_q&= \frac{-2\gamma m_e^2\pi \hbar}{\sigma k^3}\sum_\ell  \frac{\ell(\ell+1)(2\ell+1)}{\left(\hbar^2~\ell(\ell+1)\right)^2 + \gamma^2m_e^2k^2}\label{exactf}, \tag{S15}
\end{align}
where $\sigma$ is the total scattering cross-section. For a given kinetic energy range and $\ell$ value, the dominant term of the denominator in this equation is different. Let us consider that for all $\ell>\ell'$ the first term on the denominator dominates over the second one and vice versa for $1\leq\ell\leq\ell'$. With this approximation, the Wigner delay can be split into two parts as,  
\begin{equation}\tag{S16}
    \tau_q \approx \frac{-2\gamma m_e^2\pi \hbar}{\sigma k^3}\left[\sum_{\ell=1}^{\ell'}  \frac{\ell(\ell+1)(2\ell+1)}{\gamma^2m_e^2k^2} + \sum_{\ell=\ell'+1}^{\ell_{max}}  \frac{(2\ell+1)}{\left(\hbar^4~\ell(\ell+1)\right) } \right ]
\end{equation}
The total scattering cross-section $\sigma$ can be approximated to be the area of the molecule ($S_n$) itself: $\sigma \approx S_n$. With that, the expression for Wigner delay can be further simplified to, 
\begin{equation}\tag{S17}
    \tau_q \approx -\frac{B_1}{\epsilon_e^{1.5}}-\frac{B_2}{S_n^2\epsilon_e^{2.5}},
\end{equation}
where $B_1$ and $B_2$ are constants with appropriate dimensions. As can be noticed from this general formula applicable for a 2D molecular system, the Wigner delay arising from the quadrupole nature of the interaction potential depends explicitly on the area of the molecule and it is negative for all electron kinetic energies, $\epsilon_e$.

To apply this formula in case of naphthalene, we note that according to our SE-DFT calculations, for the kinetic energy window of 10-30~eV range, we have $\ell'\approx 3$ and the number of partial waves is limited to $\ell_{max}\approx 20$. For naphthalene, we have $S_n\approx0.34$ nm$^2$. The lengths for the C$=$C and C$-$H bonds in naphthalene have been measured previously using laser spectroscopic technique \cite{BabaJCP2011}. From there, we could deduce the effective radius of the naphthalene molecule to be $R\approx 6.2a_0$, with $a_0$ being the Bohr radius. We note that the value of $\mathcal{D}=-43 \times 10^{-40}$ Cm$^{2}$ calculated using such a simplistic picture matches quantitatively previous measurements \cite{Calvert1980} of the quadrupole moment in the case of neutral naphthalene molecules: $-45 \times 10^{-40}$ Cm$^{2}$. 

By plugging in all these values in Equation S16, the Wigner delay (in femtoseconds) can be expressed as a function of the electron kinetic energy, $\epsilon_e$ (in eV) and the surface area, $S_n$ (in nm$^2$) of naphthalene as follows,
\begin{equation}\tag{S18}
    \tau_q~\textrm{[fs]} \approx -\frac{2}{\epsilon_e^{1.5}}-\frac{1.25}{S_n^2\epsilon_e^{2.5}}.
    \label{SOMnaphformulea}
\end{equation}    
For naphthalene, this leads to a Wigner delay of $-64$~attoseconds at around $12.5$~eV photoelectron kinetic energy. For comparison, the use of the exact formula (Eq.~\ref{exactf}) yields  $-54$~attoseconds as the Wigner delay.

\newpage
\section*{Tables - 'Extended Data'}
\begin{table}[!htb]
    \centering
    \begin{tabular}{|c|c|c|c|c|}
    \hline
                        &  $\tau_{exp}$ (Fig. \ref{Statistics_raw})& $\tau_{cc}$ \cite{Dahlstrom2012SM}& $\tau_{2h\nu}^{Ar}$ \cite{MauritssonPRA2005SM} & $\tau_{mol}$\\
    \hline                        
    Naph sideband 14    & $6 \pm 8$    & -48       & -58        & $-4 \pm 8$   \\
    Naph sideband 16    & $-41 \pm 4$  &  -39      &   -36      & $-38 \pm 4$ \\
    Naph sideband 18    & $-11 \pm 19$  & -32       & -26        & $-4 \pm 19$  \\
    Naph sideband 20    & $1 \pm 10$   & -28       & -16        & $ 13 \pm 10$ \\
\hline
    Ada sideband 14     & $60 \pm 3$    & -52       & -58       & $54 \pm 3$   \\
    Ada sideband 16     & $-22 \pm 9$    & -42      & -36       & $-16 \pm 9$   \\    
    Ada sideband 18     & $23 \pm 21$    & -35      & -26       & $32 \pm 21$   \\    
    Ada sideband 20     & $5 \pm 6$    & -29        & -16       & $18 \pm 6$   \\
    \hline
    Pyr sideband 14     & $26 \pm 15$    & -45        & -58       & $13 \pm 15$   \\
    \hline
    Fluo sideband 14    & $13 \pm 5$      & -47        & -58       & $2 \pm 5$    \\
    \hline
    
    \end{tabular}
    \caption{Summary of the experimental measurements and the parameters used to obtain the absolute Wigner delays for naphthalene (Naph), adamantane (Ada), pyrene (Pyr) and fluorene (Fluo). The values are given in attoseconds.}
    \label{SummaryIonizTime}
\end{table}

\let\oldthebibliography=\thebibliography  
\let\oldendthebibliography=\endthebibliography
\renewenvironment{thebibliography}[1]{
    \oldthebibliography{#1}
    \setcounter{enumiv}{28}                        
}{\oldendthebibliography}

\end{document}